\documentclass[11pt]{article}

\usepackage{graphicx}
\usepackage{amssymb}
\usepackage[figuresright]{rotating}
\usepackage{subfigure}
\usepackage{amsmath}
\usepackage{hyperref}
\usepackage{multirow}
\usepackage{enumerate}
\usepackage{color}
\usepackage{tabu}
\usepackage{natbib}    
\usepackage{JASA_manu}
\usepackage{threeparttable}
\usepackage{chngcntr}
\usepackage{setspace}
\usepackage{mdwlist}
\usepackage{rotating}
\usepackage{float}
\usepackage{setspace}
\begin{document}
    \date{}

  \title{\bf \large A Simultaneous Inference Procedure to Identify Subgroups from RCTs with Survival Outcomes: Application to Analysis of AMD Progression Studies}
  \author{Yue Wei\hspace{.2cm}\\
    Department of Biostatistics, University of Pittsburgh, PA, USA\\
    email: \texttt{yuw95@pitt.edu}
    and \\
    Jason C. Hsu\\
    Department of Statistics, The Ohio State University, Columbus, OH, USA\\
    email: \texttt{jch@stat.osu.edu}
    and \\
    Wei Chen \\
    Department of Pediatrics, University of Pittsburgh \\ Children's Hospital, Pittsburgh, PA 15224, USA \\
    email: \texttt{wei.chen@chp.edu}
    and \\
    Emily Y. Chew\\
    National Eye Institute, National Institutes of Health, Bethesda, MD, USA\\
    email: \texttt{echew@nei.nih.gov}
    and \\
    Ying Ding\\
    Department of Biostatistics, University of Pittsburgh\\130 De Soto Street, Pittsburgh, PA 15261, USA\\
    email: \texttt{yingding@pitt.edu}\\
    phone: \texttt{412-624-9407}
    }
  \maketitle

\mbox{}



\newpage
\begin{abstract}
With the uptake of targeted therapies, instead of the ``one-fits-all" approach, modern randomized clinical trials (RCTs) often aim to develop treatments that target a subgroup of patients. Motivated by analyzing the Age-Related Eye Disease Study (AREDS) data, a large RCT to study the efficacy of nutritional supplements in delaying the progression of an eye disease, age-related macular degeneration (AMD), we develop a simultaneous inference procedure to identify and infer subgroups with differential treatment efficacy in RCTs with survival outcome. Specifically, we formulate the multiple testing problem through contrasts and construct their simultaneous confidence intervals, which control both within- and across- marker multiplicity appropriately. Realistic simulations are conducted using real genotype data to evaluate the method performance under various scenarios. The method is then applied to AREDS to assess the efficacy of antioxidants and zinc combination in delaying AMD progression. Multiple gene regions including \textit{ESRRB-VASH1} on chromosome 14 have been identified with subgroups showing differential efficacy. We further validate our findings in an independent subsequent RCT, AREDS2, by discovering consistent differential treatment responses in the targeted and non-targeted subgroups been identified from AREDS. This simultaneous inference approach provides a step forward to confidently identify and infer subgroups in modern drug development.
\end{abstract}

\vspace*{.3in}

\noindent%

\newpage
\section{Introduction}
\label{sec:intro}
With rapid advances in the understanding of human diseases, the paradigm of medicine shifts from ``one-fits-all” to ``targeted therapies” or ``precision medicine". One aspect of precision medicine research is to develop new therapies that target a subgroup of patients. The subgroups are usually defined by ``markers'', where the markers could be genotype information (such as mutation of a certain gene), expression level of certain protein(s), disease severity or any biologically plausible factors. For example, the $KRAS$ status of patients with metastatic colorectal cancer is associated with the progression free survival when treated by panitumumab monotherapy \citep{KRAS}, and the vemurafenib (Zelboraf) was approved for treating the $BRAF$-mutated metastatic melanoma \citep{BRAF}. For another example, multiple genetic variations including $FCER2$ have been reported to be associated with the drug response among asthma patients using inhaled corticosteroids \citep{FCER2}.

The drug development process typically involves comparing a new treatment ($Rx$) with a control ($C$, such as a standard-of-care) through randomized clinical trials (RCTs), and treatment efficacy is the ``relative'' effect between $Rx$ and $C$. Finding the subgroups of patients that exhibit enhanced treatment efficacy is a problem at the heart of modern drug development. To confidently identify subgroups, it is often necessary to infer treatment efficacy in each group and some combination of groups. For example, for a single nucleotide polymorphism (SNP) that separates patients into three groups (denoted by $AA$, $Aa$ and $aa$), one may have to decide whether the new treatment should target a single group (e.g., $aa$) or a combination of groups (e.g., $\{Aa, aa\}$). In this case, the treatment efficacy in both single genetic groups and their combinations need to be assessed. As shown by \cite{binary2019}, when population is a mixture of subgroups with heterogeneous efficacy, the efficacy measure needs to respect the logic relationship among subgroups and their combinations. Take the SNP case as an example, if the treatment efficacy for $Aa$ is $x$, and for $aa$ is $y$, then the efficacy for $\{Aa, aa\}$ should be within $[x, y]$ (assuming $x \le y$). This may sound trivial. However, several commonly used efficacy measures such as the hazard ratio (for time-to-event data) and the odds ratio (for binary data) do not satisfy this relationship \citep{SME, binary2019}. Note that this logic-respecting is a property of an efficacy measure, not a property of the type of data that quantifies the outcome, nor a statistical model for analyzing such data.

Our research is highly motivated from analyzing the Age-Related Eye Disease Study (AREDS) data, a RCT to study the risk factors for the age-related macular degeneration (AMD) and assess the effect of micronutrients on delaying the progression to late AMD \citep{AREDS}. AMD is a polygenic and progressive neurodegenerative disease, which is a leading cause of blindness in elderly. Patients can progress to one or both forms of late-AMD -- central geographic atrophy (GA) and choroidal neovascularization (CNV). The AREDS study collected DNA samples of consenting participants and performed genome-wide genotyping \citep{AMD_genetic_2016}. Many genetic studies have shown that the development or the progression of AMD is associated with various genetic risk factors \citep{Seddon_2007, AMD_genetic_2016, Yan2017_AMD, Tao_interval}. Specifically, in two recent genomewide association studies for AMD progression using the AREDS data \citep{Yan2017_AMD, Tao_interval}, where time-to-late-AMD is the outcome, multiple variants from \textit{ARMS2-HTRA1} and \textit{CFH} gene regions have been discovered to be associated with AMD progression. Besides association analyses where no treatment is involved, multiple research groups also investigated whether variants from these two gene regions are associated with differential treatment responses. A recent review article by \cite{PrecisionMed_2018} summarized the controversial findings. Research groups such as \cite{Klein_2008} and \cite{Seddon2016_AREDS} reported that  genetic variants from \textit{CFH} and \textit{ARMS2} regions were found to be associated with differential responses to the antioxidants plus zinc treatment. However, the AREDS investigators reported no significant associations between \textit{CFH} and \textit{ARMS2} regions and the nutritional supplements, when multiplicity adjustment has been taken into account \nocite{Emily_2015}. To fully understand the effects of those nutritional supplements on AMD progression and to infer whether there are genetic subgroups with enhanced treatment efficacy, a rigorous statistical procedure that can simultaneously identify and infer subgroups for time-to-event outcome is required.

There are existing methods for detecting heterogeneous treatment effects across groups for time-to-event outcome. One simple but broadly used method is to test the treatment-by-marker interaction in the Cox Proportional Hazards (CoxPH) model \citep{Cox}. However, this method cannot provide which group to target directly, nor can it provide inference on subgroup-specific efficacy. The second type of approach focuses on testing the existence of a subgroup (with an enhanced treatment effect) using either a logistic-Cox model for the response in each subgroup and the latent subgroup membership \citep{logisticCox} or a new CoxPH model including a nonparametric component for the covariate in the control group and a subgroup-treatment-interaction effect defined by a change plane \citep{nonparametricTest}. Similar to the interaction test, additional steps are needed to provide inference in the targeted and non-targeted group. The third common approach utilizes tree-based regression models, including RECursive Partition \citep{RECPAM, Negassa2005}, interaction trees \citep{interactiontrees}, SIDES \citep{SIDES}, GUIDE \citep{GUIDE} and etc. One appealing feature of using these methods is that it can provide the tree structure which incorporates multiple covariates at the same time. However, none of these methods provides inference for treatment efficacy in both targeted group and non-targeted group, and some approaches use illogical efficacy measures. The targeted treatment development process involves the co-development of a drug compound and a companion diagnostic tool that identifies the suitable subgroup of patients for the drug to target. Therefore, the subgroup has to be ``simple" (usually defined by one or two biomarkers) for clinical and regulatory feasibility. In this article, we develop a multiple-testing-based approach which aims to simultaneous identify and infer ``simple'' subgroups with enhanced treatment efficacy.

The article is organized as follows. Section \ref{sec:meth} introduces the logic-respecting efficacy measure for time-to-event outcomes that we choose to use and its associated properties, and with that efficacy measure how we formulate the contrasts to identify subgroups and adjust for the multiplicity. Section \ref{sec:sim} presents simulations to show finite sample performance of the proposed method and uses realistic simulations to summarize practical rules for the use of the method. Then we apply our method on the Age-related Eye Disease Study (AREDS) data and present our findings in Section \ref{sec:application}. Finally, we discuss and conclude in Section \ref{sec:conclusion}.

\section{Methods}
\label{sec:meth}
In this article, we deal with ordinal categorical markers which separate the population into a few groups (e.g., three groups by a SNP, or four groups by the immunohistochemistry test). Below we use the scenario of ``three-category" marker $(M=0,1,2)$ to illustrate our method. Brief discussions are provided in Section \ref{sec:conclusion} regarding how to generalize the method to handle markers with more categories or continuous markers.

\subsection{Issues with hazard ratio}
\label{sec:hr}
First, we demonstrate why the commonly used HR is not a suitable efficacy measure when the population is a mixture of subgroups. Let $\mathrm{HR}_0, \mathrm{HR}_1 , \mathrm{HR}_2 $ denote the hazard ratio for each subgroup defined by $M$, and let $ \mathrm{HR}_{01} $ denote the HR for the $\{M=0,1\}$ combined. HR is not logic-respecting in the sense that even if both $\mathrm{HR}_{0}$ and $\mathrm{HR}_{1}$ are constant, the combined population typically does not have a constant HR. In fact, the HR of the mixture population is usually a complex function of time, with values at some time points outside of $[\mathrm{HR}_{0}, \mathrm{HR}_{1}]$. This is because $\mathrm{HR}_{01}$ can not be expressed as a weighted combination of $\mathrm{HR}_{0}$ and $\mathrm{HR}_{1}$. The combination can only be made on density or cumulative density functions, not on the hazard ratio scale. For example, we can generate data from a Weibull distribution where $\mathrm{HR}_{0}=\mathrm{HR}_{1}=\exp(-0.2)=0.82$, with an equal prevalence of the two subgroups. However, the true $\mathrm{HR}_{01}$ is a smooth function of time and goes above 0.82 when $t$ is large. In this case, it is possible that another subpopulation has $\mathrm{HR}_2 = 0.84$ and then at some large time (e.g., $t>1.1$), $\mathrm{HR}_0 = \mathrm{HR}_1 (=0.82) < \mathrm{HR}_2 (=0.84) < \mathrm{HR}_{01} (= 0.85) < 1 $, meaning $\{M=2\}$ exhibits more efficacy than $\{M=0,1\}$ \emph{combined}, but does not exhibit more efficacy than either $\{M=0\}$ or $\{M=1\}$. Thus using HR as the efficacy measure can lead to paradoxical findings in patient targeting.

\subsection{Ratio of quantile survival times and its property}
\label{sec:logic}
Realizing that HR is not suitable to use in the presence of mixture population, a different measure needs to be considered. \cite{SME} demonstrated that the ratio or difference of mean or median survival times (between $Rx$ and $C$) is logic-respecting by guaranteeing $\mu_0 \le \mu_{01} \le \mu_1$, where $\mu$ denotes the efficacy measure. In this manuscript, we choose ratio of quantile survival times as our efficacy measure and demonstrate that it has a unique property under the CoxPH model that we consider. Assume the time-to-event data fit the following model:
\begin{eqnarray}
\label{CoxPH}
h(t|Trt,M,X) = h_{0}(t)\{\beta_{1}I(Trt=1)+\beta_{2}I(M=1)+\beta_{3}I(M=2)+ \nonumber \\ \beta_{4}I(Trt=1) \times I(M=1)+\beta_{5}I(Trt=1) \times I(M=2)+\beta_{6}\mathbf{X}\},
\end{eqnarray}
where $Trt = 0$ $(C)$ or $1$ $(Rx)$ is the treatment assignment, $M=0$, $1$, or $2$ is the marker we consider for testing, and $h_{0}(t)=h(t|C,M=0,\mathbf{X}=\mathbf{0})$ is the hazard function for the $\{M=0\}$ group receiving $C$, with additional covariates set to 0. Further we assume that the baseline survival function is from a Weibull distribution with scale $\lambda$ and shape $k$, which corresponds to $h_{0}(t)=h(t|C,M=0,\mathbf{X}=\mathbf{0})=\frac{kt^{k-1}}{\lambda^{k}}$. Denote by $\nu^{Trt}_{M, \tau}$ the corresponding quantile ($\tau$) survival time in each marker-by-treatment group. Let $\theta_{l}=e^{\beta_{l}}, l=1,\dots,5$ and $\theta_{6}=e^{\mathbf{\beta_{6}X}}$ with $\beta$s from (\ref{CoxPH}), then by setting the survival function for each group equal to $\tau$, the corresponding survival and their ratios can be directly calculated as follows:
\begin{eqnarray}
\label{medsurv}
\nu^{Rx}_{0,\tau}=\lambda(\frac{\log{\tau}}{\theta_{1}\mathbf{\theta_{6}}})^{\frac{1}{k}}, \quad \nu^{C}_{0,\tau}=\lambda(\frac{\log{\tau}}{\mathbf{\theta_{6}}})^{\frac{1}{k}},\quad r_{0}=\frac{\nu^{Rx}_{0,\tau}}{\nu^{C}_{0,\tau}}=(\frac{1}{\theta_{1}})^{\frac{1}{k}}, \nonumber \\
\nu^{Rx}_{1,\tau}=\lambda(\frac{\log{\tau}}{\theta_{1}\theta_{2}\theta_{4}\mathbf{\theta_{6}}})^{\frac{1}{k}}, \quad  \nu^{C}_{1,\tau}=\lambda(\frac{\log{\tau}}{\theta_{2}\mathbf{\theta_{6}}})^{\frac{1}{k}},\quad  r_{1}=\frac{\nu^{Rx}_{1,\tau}}{\nu^{C}_{1,\tau}}=(\frac{1}{\theta_{1}\theta_{4}})^{\frac{1}{k}}, \nonumber \\
\nu^{Rx}_{2,\tau}=\lambda(\frac{\log{\tau}}{\theta_{1}\theta_{3}\theta_{5}\mathbf{\theta_{6}}})^{\frac{1}{k}}, \quad \nu^{C}_{2,\tau}=\lambda(\frac{\log{\tau}}{\theta_{3}\mathbf{\theta_{6}}})^{\frac{1}{k}},\quad r_{2}=\frac{\nu^{Rx}_{2,\tau}}{\nu^{C}_{2,\tau}}=(\frac{1}{\theta_{1}\theta_{5}})^{\frac{1}{k}}. \nonumber
\end{eqnarray}

\indent It can be seen that the quantile survival time for each group depends on the baseline characteristics ($\mathbf{\theta_{6}})$, but the ratio does \textbf{not}. We name it as the \textit{covariate invariant} property. This property is unique to this efficacy measure, which is attractive as it makes the comparison (between $Rx$ and $C$) simple. Further it can be shown that this property also holds in the combined groups. For example, suppose we are interested in the ratio of quantile survival times in the mixture population of $\{M=0, 1\}$ (denoted as $r_{01}^{\tau}$). We can calculate $r_{01}^{\tau}$ from its definition, $r_{01}^{\tau} =\frac{\nu^{Rx}_{01,\tau}}{\nu^{C}_{01,\tau}}$, where $\nu^{Rx}_{01,\tau}$ and $\nu^{C}_{01,\tau}$ can be obtained by solving the following equations,
\begin{eqnarray}
\label{combinedsurv}
t=\nu^{Rx}_{01,\tau}&:& p_{0}e^{\{-\theta_{1}\mathbf{\theta_{6}}(\frac{t}{\lambda})^{k}\}}+(1-p_{0})e^{\{-\theta_{1}\theta_{2}\theta_{4}\mathbf{\theta_{6}}(\frac{t}{\lambda})^{k}\}}=\tau,\nonumber\\
t=\nu^{C}_{01,\tau}&:& p_{0}e^{\{-\mathbf{\theta_{6}}(\frac{t}{\lambda})^{k}\}}+(1-p_{0})e^{\{-\theta_{2}\mathbf{\theta_{6}}(\frac{t}{\lambda})^{k}\}}=\tau,
\end{eqnarray}
with $p_{0}$ representing the prevalence of $M=0$ in the combined population $\{0,1\}$. By combining the two groups at the probability level, this calculation follows the subgroup mixable estimation (SME) principle \citep{SME}. Let $x^{Rx}_{01,\tau}=\mathbf{\theta_{6}}(\frac{\nu^{Rx}_{01,\tau}}{\lambda})^{k}$ and $x^{C}_{01,\tau}=\mathbf{\theta_{6}}(\frac{\nu^{C}_{01,\tau}}{\lambda})^{k}$, then we have $r_{01}^{\tau}=\frac{\nu^{Rx}_{01,\tau}}{\nu^{C}_{01,\tau}}=(\frac{x^{Rx}_{01,\tau}}{x^{C}_{01,\tau}})^{\frac{1}{k}}$. Since the solutions for $x^{Rx}_{01,\tau}$ and $x^{C}_{01,\tau}$ from equation (\ref{combinedsurv}) are free of $\mathbf{\theta_{6}}$, we also have the covariate-invariant property for $r_{01}^{\tau}$.

\subsection{Confident Effect 4 contrasts (CE4) for ratio of quantile survival times}
\label{sec:CE4}
In targeted therapy development, researchers are interested in (1) whether there exists a subgroup with enhanced treatment efficacy and (2) the treatment efficacy in both targeted and non-targeted subgroups (for appropriate drug labeling) \citep{binary2019}. To answer both questions simultaneously, we propose to use contrasts to compare efficacy between different subgroups and combination of subgroups. Under the same scenario with three groups defined by a marker ($M=0$, $1$, or $2$), in order to get a complete ordering of the treatment efficacy in all possible groups, we propose the following four contrasts:
\begin{eqnarray}
\label{CE4}
\log\kappa_{(1,2):0}=\log(\frac{r_{12}}{r_{0}})=\log r_{12}-\log r_{0}, \quad \log\kappa_{1:0}=\log(\frac{r_{1}}{r_{0}})=\log r_{1}-\log r_{0},\nonumber\\
\log\kappa_{2:(0,1)}=\log(\frac{r_{2}}{r_{01}})=\log r_{2}-\log r_{01}, \quad \log\kappa_{2:1}=\log(\frac{r_{2}}{r_{1}})=\log r_{2}-\log r_{1}.
\end{eqnarray}
We drop $\tau$ in the notation as $\tau$ is pre-specified. Moreover, these contrasts are built on the log scale of the efficacy measure since previous experience demonstrates the normality approximation seems to work better on the log scale (as compared to the original scale) \citep{SME}. In fact these four contrasts are analogous to the contrasts proposed in \cite{ding2018} where the efficacy in their case is measured by a continuous outcome. As pointed by \cite{ding2018}, they are equivalent to testing the following eight one-sided null hypotheses where each one is to test an inequality against its complement (i.e., $H_0: \kappa_{(1,2):0}\leq 1$ vs $H_a: \kappa_{(1,2):0} >1$) rather than testing a zero null (such as $H_0: \kappa_{(1,2):0}=1$).
\begin{gather*}
\label{8tests}
H^{\leq}_{(1,2):0}: \kappa_{(1,2):0}\leq 1, \ H^{\leq}_{(0,1):2}: \kappa_{(0,1):2}\leq 1, \ H^{\leq}_{1:0}: \kappa_{1:0}\leq 1, \ H^{\leq}_{1:2}: \kappa_{1:2}\leq 1\\
H^{\leq}_{2:(0,1)}: \kappa_{2:(0,1)}\leq 1, \ H^{\leq}_{0:(1,2)}: \kappa_{0:(1,2)}\leq 1, \ H^{\leq}_{2:1}: \kappa_{2:1}\leq 1, \ H^{\leq}_{0:1}: \kappa_{0:1}\leq 1.
\end{gather*}
From these eight one-sided tests, we are able to tell which subgroup or combination of subgroups exhibits a differential efficacy than its complementary group.

We propose to use simultaneous confidence intervals on these four contrasts so that we can identify differential subgroup(s) and infer their efficacy simultaneously. Note that level $ 100(1-\alpha)\% $ simultaneous confidence intervals for those contrasts effectively form a level-$ \alpha $ interaction test: reject the null hypothesis of no interaction between Treatment effect ($ Trt $) and marker group ($ M $) if at least one of the confidence intervals does not contain zero. Moreover, this formulation of assessing ``interaction" effect is advantageous toward patient targeting as it allows decision-making based on clinically meaningful differences (reflected from confidence intervals on efficacy comparisons) instead of a mere statistical significance (such as the $p$-value from the typical interaction test).
To estimate the four contrasts from (\ref{CE4}) under our model (\ref{CoxPH}), we propose the following three steps and name this approach as ``CE4-Weibull":
\begin{enumerate}[-]
\item Step 1. Estimate all the parameters in the Weibull model (e.g., $\lambda, k, \beta_{l}, l=1,\dots,6$).
\item Step 2. Estimate the quantile survival times $\nu^{Trt}_{0,\tau}$, $\nu^{Trt}_{1,\tau}$, $\nu^{Trt}_{2,\tau}$, $\nu^{Trt}_{01,\tau}$, $\nu^{Trt}_{12,\tau}$, $Trt=Rx$ or $C$, based on the parameter estimates obtained in Step 1 and their associated variance covariance matrix.
\item Step 3. Estimate $r_{0}, r_{1}, r_{2}, r_{01}$ and $r_{12}$ and their variance covariance using estimators obtained from Step 2 and then calculate the four contrasts CE4.
\end{enumerate}

The estimated variance covariance matrices in Step 2 and Step 3 can be obtained using the Delta method. Note that in Step 2, the Delta method for implicitly defined random variables \citep{eDelta} needs to be applied since the quantile survival times in the combined groups (e.g., $\{0,1\}$ and $\{1,2\}$) are not explicitly defined, but rather from solving equations like (\ref{combinedsurv}).

The estimated CE4 will asymptotically follow a multivariate normal distribution and the simultaneous confidence intervals can be then derived as follows. We compute the quantile $q$ such that the four simultaneous confidence intervals
$$\log(\hat{\kappa}_{g})-q\hat{s}_{gg}<\log(\kappa_{g})<\log(\hat{\kappa}_{g})+q\hat{s}_{gg}, \ g=\{(1,2):0, \ 2:(0,1), \ 1:0, \ 2:1\}$$
have a coverage probability $1-\alpha$, that is, the joint probability
$$Pr\Big(\frac{|\hat{\log(\kappa}_{g})-\log(\kappa_{g})|}{\hat{s}_{gg}}<q, \ g=\{(1,2):0, \ 2:(0,1), \ 1:0, \ 2:1\}\Big)=1-\alpha,$$ where $\hat{s}^2_{gg}$ is the variance estimator for $\log(\hat{\kappa}_{g})$. The R package \{mvtnorm\} can be used to obtain this $q$ value. Then the $p$-value (from testing the eight one-sided null hypotheses simultaneously) can be obtained from the multivariate normal distribution as well, which corresponds to the smallest $p$-values from the four single contrasts. If any of the four contrasts does not cover 0, it suggests that there exists subgroup(s) with differential treatment efficacy.

\subsection{Multiplicity adjustment across biomarkers}
In targeted treatment development, typically a large collection of markers need to be tested in order to identify subgroups. Therefore, there are two families of inferences need to be considered: within a marker and across markers. Specifically,  strong control of familywise error rate (FWER) for inference within a marker is desired, since the consequence of an incorrect inference may target a wrong subgroup, which is serious. The simultaneous confidence intervals obtained from our CE4-Weibull method appropriately controls the within-marker FWER. While the error rate for inference across multiple markers can be controlled less stringently, since multiple candidate markers can be identified for tailoring (which may indicate largely overlapped subgroups to target), and therefore the $per family$ error rate seems acceptable.

Suppose there are a total of $K$ markers to be tested. Denote by $V_{k}$ the number of confidence intervals that fail to cover the true values for the $k^{th}$ marker. Then the FWER for the $k^{th}$ SNP is $\alpha_{k}=P\{V_{k}>0\}=E[I_{\{V_{k}>0\}}]$. For inference across SNPs, denote by $V_{\ast}$ the number of markers that have at least one of its confidence intervals failing to cover its true value. Then the $per family$ error rate is $E[V_{\ast}]=E\left[\sum\limits_{k=1}^{K}I_{\{V_{k}>0\}}\right]=\sum\limits_{k=1}^{K}P\{V_{k}>0\}=\sum\limits_{k=1}^{K}\alpha_{k}$. We suggest to use the simple \textit{additive} adjustment as proposed by \cite{ding2018}, of which by setting the desired $per family$ error rate as $m$ (a pre-specified positive integer), the familywise $\alpha_{k}$ for each marker is then set to be $\frac{m}{K}$ (same across all markers).

When SNPs are the biomarkers to define subgroups, the screening process seems similar to a genome-wide association study (GWAS). However, our proposal controls for $per family$ error rate instead of the commonly used false discovery rate (FDR). In GWAS, it is plausible \emph{biologically} that the vast majority of the SNPs are not associated with the specific disease. However, when treatments are involved, the biological processes become more complex, and zero-nulls of no-difference (e.g., phrased as $H_0: \kappa_{(1,2):0}=1$) are statistically false, which was first observed in the setting of \cite{ding2018}, where the treatment efficacy was simulated based on a single causal SNP with no random error being added. It was found that practically all other SNPs would appear ``associated'' with the outcome (as sample size reaches infinity) when analyzed in a SNP-by-SNP fashion. The reason is that most SNPs are not ``orthogonal'' to each other, and thus any SNP will appear somewhat associated with treatment outcome as long as the distribution for proportions of being $\{AA, Aa, aa\}$ in this SNP and the causal SNP are not independent, which is most of the cases. When there are no zero-nulls statistically, the ``false" discovery seems lame, and the $per family$ rate is preferred by providing more meaningful candidates. We have more discussions about the zero-null issue in Section \ref{sec:sim}.

\section{Realistic Simulations}
\label{sec:sim}
\subsection{Single SNP simulations}
We use SNPs as the biomarkers in all simulation studies. First, simulation studies based on a single SNP were conducted to investigate the finite sample performance of the proposed CE4-Weibull method. We considered three scenarios: (1) No SNP effect, i.e., \textit{Rx} is \textit{not} efficacious for any genotype group; (2) The allele \textit{a} has a dominant beneficial effect on \textit{Rx}; (3) The allele \textit{a} has a recessive beneficial effect on \textit{Rx}. The SNP was simulated from a multinomial distribution with $(P_{aa}=0.16, P_{Aa}=0.48, P_{AA}=0.36)$ (corresponding to minor allele frequency (MAF) of 0.4). Survival times were simulated from model (\ref{CoxPH}) with scale $\lambda = 2 $ and shape $k = 1.25$. The parameters $(\beta_{1},\beta_{2},\beta_{3},\beta_{4},\beta_{5},\beta_{6})$ were set to be $(0, -0.8, -0.8, 0, 0, 0)$, $(0, -0.8, -0.8,-0.6, -0.6, 0)$ and $(0, -0.8, -0.8, 0, -0.6, 0)$ for the three scenarios respectively. The censoring times were generated from an independent uniform distribution $U(a,b)$ with \textit{a} and \textit{b} chosen to yield 20\% and 50\% censoring rates. We chose the quantile $\tau$ as $0.5$ which corresponds to the median survival. The true values of the CE4 contrasts $(\kappa_{(1,2):0}, \kappa_{2:(0,1)}, \kappa_{1:0}, \kappa_{2:1})$ using the ratio of median survival as the efficacy measure for each scenario are: (1) $(1, 1, 1, 1)$, (2) $(1.62, 1.27, 1.62, 1)$, and (3) $(1.12, 1.62, 1, 1.62)$.

\begin{figure}
\begin{center}
\includegraphics[width=6in]{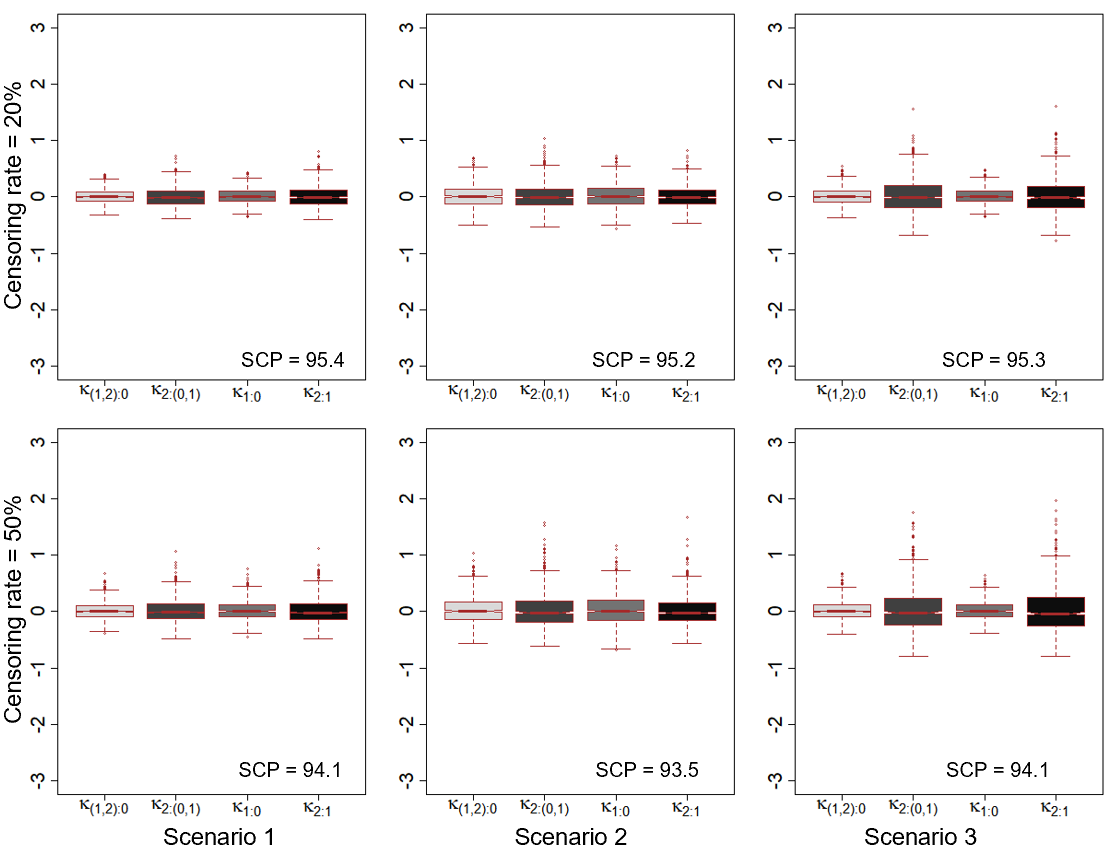}
\end{center}
\caption{Finite sample performance of CE4-Weibull: bar plots of the biases of CE4 estimators. \label{fig:simulation}}
\end{figure}

We ran 1000 simulations for each scenario, with sample size 500 for each treatment arm and the results are summarized in Figure~\ref{fig:simulation}.
Across all the scenarios, the biases of the CE4 estimates are minimal and the coverage probabilities for the simultaneous confidence intervals (SCP) are all close to 95$\%$. Larger variations are observed in biases of $\hat{\kappa}_{2:(0,1)}$ and $\hat{\kappa}_{2:1}$, especially under scenario 3. This is because under the recessive effect setting, $Rx$ is only efficacious in $\{aa\}$ patients, which is a small proportion of the total population ($16\%$). Therefore, the contrasts involving the comparison between ${aa}$ and other group (or the combination of other groups) have larger variances.

\subsection{Realistic simulations}
\label{sec:realsim}
To understand the performance of the proposed CE4-Weibull method in the real genetic setting with a large number of SNPs, we used the chromosome-wide data from AREDS. Among those ($>2700$) participants who had DNA collected and genotyped, we randomly selected 1000 Caucasian participants and randomly ``assigned" them in a 1:1 ratio to the new treatment \textit{Rx} and a standard care \textit{C}. Since AMD is known as a polygenic disorder and many studies have discovered or confirmed genetic risk variants associated with AMD, in this realistic simulation, we selected a variant $rs2284665$ from the well-known AMD risk gene region $ARMS2$ on Chromosome 10 as the causal SNP, and assumed the minor allele of this variant has a dominant beneficial effect on $Rx$. We kept the three genotype groups (defined by the causal SNP) balanced between $Rx$ and $C$. Similar as in the single SNP simulations, the progression times were simulated from the Weibull model with $\lambda = 2 $ and $k = 1.25$. The parameters for $(\beta_{1},\beta_{2},\beta_{3},\beta_{4},\beta_{5},\beta_{6})$ were set to be $(0, -0.5, -0.5, -0.8, -0.8, 0)$, corresponding to $(\kappa_{(1,2):0}, \kappa_{2:(0,1)}, \kappa_{1:0}, \kappa_{2:1})=(1.90, 1.51, 1.90, 1)$. The censoring rate was set to be 25\%. Therefore, the SNP data were real (from real subjects) and the progression times and SNP effects were simulated.

\begin{figure}
\begin{center}
\includegraphics[width=4.7 in]{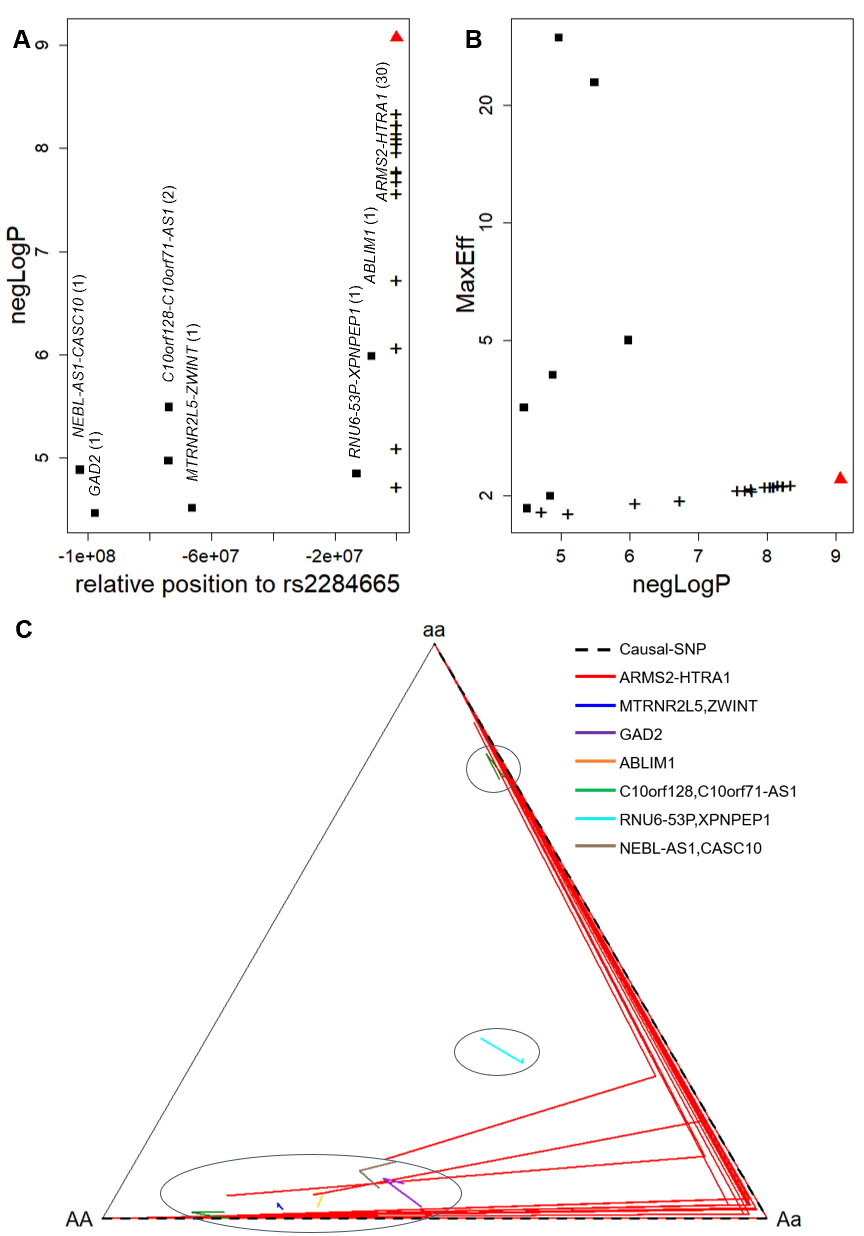}
\end{center}
\caption{37 identified SNPs from one chromosome-wide realistic simulation. \textbf{A}: -log$_{10}$(\textit{p.}CE4) $vs.$ relative position to the causal SNP; \textbf{B}: the maximum effect among CE4 $vs.$ -log$_{10}$(\textit{p.}CE4); the $\triangle$ is the causal SNP $rs2284665$ and `+'s are the SNPs that from the same rigion with the causal SNP. The rests are from other gene regions; \textbf{C}: SNP cross-talk plot.  \label{fig:onesim}}
\end{figure}

We analyzed chromosome 10 using our CE4-Weibull model and filtered the SNPs with less than three patients in each genotype group within each treatment arm, which resulted in a total of 268,053 SNPs. We set $m=10$, allowing on average 10 out of $\sim$ 270,000 SNPs with at least one confidence interval failing to cover its true value, which is equivalent to setting the $\alpha_K$ level at $3.73 \times 10^{-5}$ ($= \frac{m}{K} = \frac{10}{268,053}$). A total of 37 SNPs were identified by CE4-Weibull. Among those 37 SNPs, 30 of them are from the \textit{ARMS2-HTRA1} region, including the causal SNP. Other seven SNPs belong to six different gene regions, which are distance away from the causal gene region. Figure \ref{fig:onesim}A plotted the positions of these SNPs relative to the causal SNP, with $y$-axis ($-\log_{10}(p$.CE4)) showing the significance level of each SNP. Figure \ref{fig:onesim}B plotted MaxEff vs $-\log_{10}(p$.CE4), where MaxEff (maximal effect) is defined as the maximum absolute value among the estimated CE4 contrasts that do not cover zero. The causal SNP has the smallest $p$-value ($=8.52 \times 10^{-10}$) and with MaxEff of 2.20. Note that some top SNPs have very large MaxEff values. For example, SNP $rs10857454$ from the \textit{C10orf128-C10orf71-AS1} region has the largest MaxEff of 29.7, while its \textit{p}-value is relatively large ($=3.21 \times 10^{-6}$, close to the threshold). We caution against the situation when a huge effect size is seen, since such a huge effect for treatment efficacy is clinically unlikely. For this specific SNP, it is not surprising to see the corresponding confidence interval for $\kappa_{2:(0,1)}$ is very wide and the effective patient population only consists $1.5\%$ of the total population.

To further investigate the relationship between the top SNPs and the causal SNP, we proposed a novel SNP \textit{\textbf{cross-talk}} plot. It is based on a ternary diagram using barycentric coordinates to display the proportion of three variables that sum to one. Specifically, we projected the percentages of the $AA$, $Aa$, and $aa$ categories of the causal SNP $rs2284665$ in each of these categories of a given top SNP onto the triangular diagram, and connect the points with lines. If the SNP is highly correlated with the causal SNP in terms of the distribution of \textit{AA}, \textit{Aa}, and \textit{aa}, the percentages will be close to (1,0,0), (0,1,0) and (0,0,1), and thus the connected line segments will be long and lie closely to the two edges of the triangle. Otherwise, the three dots will be close to each other to give a short angle. For example, in Figure \ref{fig:onesim}C, the causal SNP has a perfect match in terms of the percentages with itself so the three points are the vertexes of the triangle, which makes the connected line segment coincide with the edges $AA-Aa$ and $Aa-aa$ (denoted by the dashed lines). From the plot, all 30 SNPs from \textit{ARMS2-HTRA1} region are highly correlated with the causal SNP, indicated by the long red line segments, which explains why they have been identified by CE4-Weibull. For the 7 SNPs from other regions, their line segments are all short, indicating they might have been identified due to randomness. Then we repeated this chromosome-wide realistic simulation for 100 times.


\subsection{Results from 100 realistic simulations}
\label{sec:100run}
\begin{figure}
\begin{center}
\includegraphics[width=5.5in]{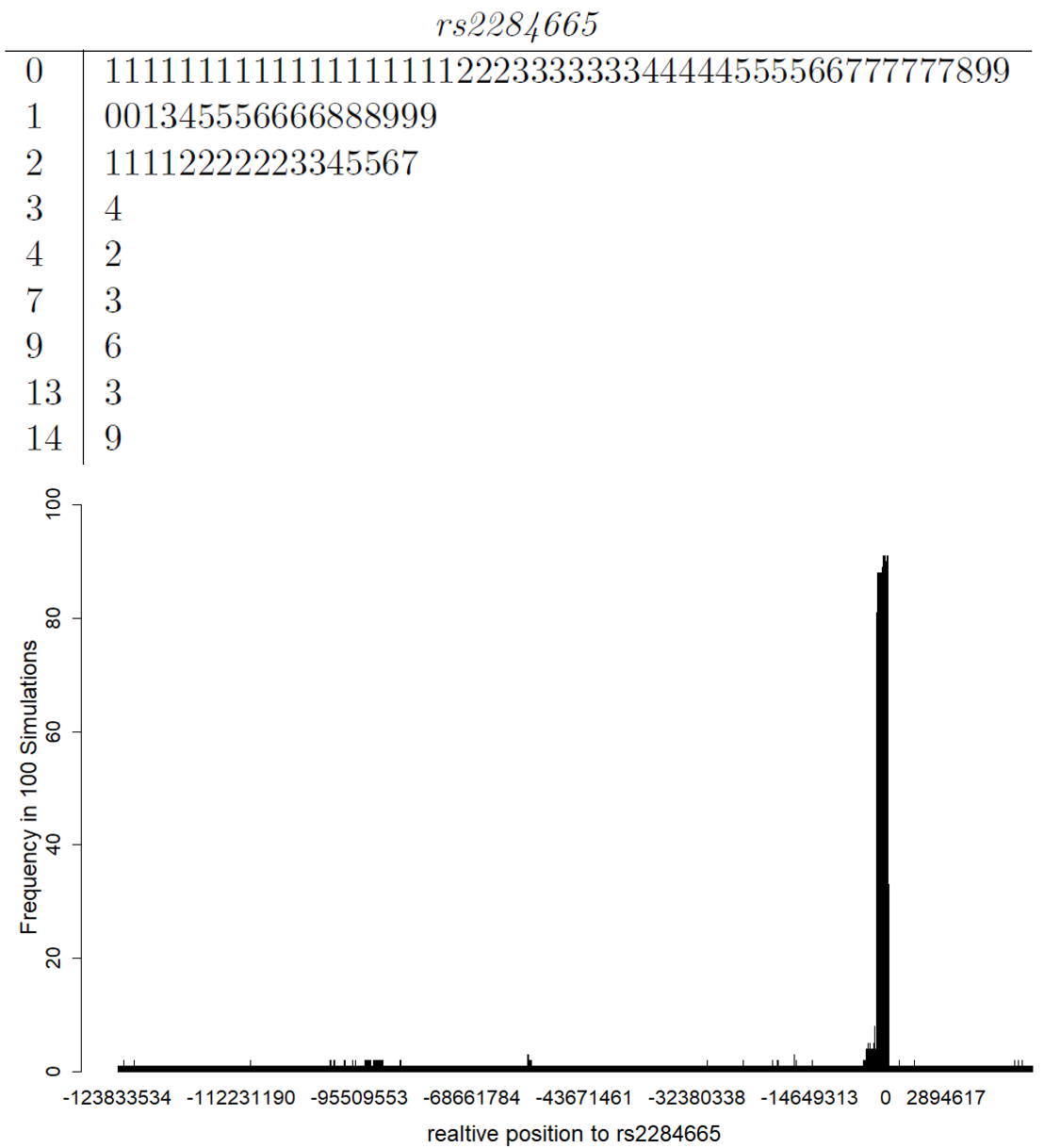}
\end{center}
\caption{\textbf{Upper:} stem-and-Leaf plot for the distribution of the ranks of the Causal SNP; \textbf{Lower:} present frequency of the identified SNPs in 100 simulations \label{fig:freq100}}
\end{figure}

In these 100 repeated runs, the SNP data are all the same but the progression times and censoring times are different due to randomness from the model. By setting $m=10$, on average there are 61 SNPs identified per run with a total of 3292 SNPs being picked at least once. The causal SNP were picked 90 out of 100 times and the distribution of the ranks is shown in the stem-and-leaf plot (upper panel in Figure \ref{fig:freq100}). Note that 84 out of 90 times the rank of the causal SNP was among top 30 and 52 times it was among top 10, indicating that our CE4-Weibull is robust in identifying the true causal SNP. The lower panel in Figure \ref{fig:freq100} summarizes all the identified SNPs from all 100 runs in terms of their relative position to the causal SNP and their frequencies of being picked up. We found that 98.9\% of the 3292 SNPs were only picked less than 5 times, which are highly likely due to randomness. While for SNPs close to the causal SNP and located in the same \textit{ARMS2-HTRA1} gene region, the probability of being selected is much higher, among which 27 SNPs were identified for more than 80\% of the times. From this repeated chromosome-wide simulations, we confirmed that there are possibilities that some SNPs are picked by random error but the true causal SNP and its surrounding SNPs can be identified with very high probabilities by CE4-Weibull. Moreover, due to the existence of linkage disequilibrium among SNPs, it is very unlikely that an isolated SNP will be the true causal SNP.

Based on the observations from our realistic simulations, we recommend the following rules to guide the selection of ``candidate" SNPs from those identified by CE4-Weibull. (1) There are multiple SNPs (e.g.,$\geq 4$) being picked from the same gene region; (2) The MaxEff should not be unrealistically large; and (3) The targeted group should be a reasonable proportion (not too small or large, e.g., $5\%-95\%$) of the total population.


\section{Application to AREDS Data}
\label{sec:application}

\subsection{Background}
AREDS is a large multi-center RCT sponsored by the National Eye Institute to evaluate the effect of antioxidants and/or zinc on delaying the progression of AMD \citep{AREDS}. The original study includes four treatment arms: placebo, antioxidants, zinc and the combination of antioxidants and zinc, where the last treatment then becomes the ``AREDS formula" dietary supplements which are now available in various drug stores. However, the treatment effects of the non-placebo arms on delaying the late AMD progression are not statistically significant \citep{prediction_genetics}. Among the four arms, we specifically investigated participants in the placebo arm ($C$) and the combination of antioxidants and zinc treatment arm ($Rx$), which included 1,170 Caucasian participants with both eyes free of advanced AMD progression when entering the study. The outcome is the time-to-late-AMD from the first progressed eye, where late-AMD is defined as the severity score reaches 9 or above (9=GA, 10=central GA, 11=CNV, 12=central GA and CNV). As shown in Table \ref{tab:tableone}, age, sex and smoking status do not differ between the two treatment arms. However, the baseline AMD severity score is significantly higher among patients who were randomized to the $Rx$ group, as compared to patients who were randomized to placebo ($4.0 \pm 2.2$ vs $2.6 \pm 2.1$). This is as expected and is due to the randomization design: patients free of AMD at baseline can only be randomized to the placebo or antioxidants arms, but not the combination arm, and thus the baseline severity score needs to be adjusted in the analysis. The overall censoring rate is about 75\%, so we used $\tau=0.75$ as our quantile.

\subsection{CE4-Weibull on AREDS}
\begin{figure}
\begin{center}
\includegraphics[width=6in]{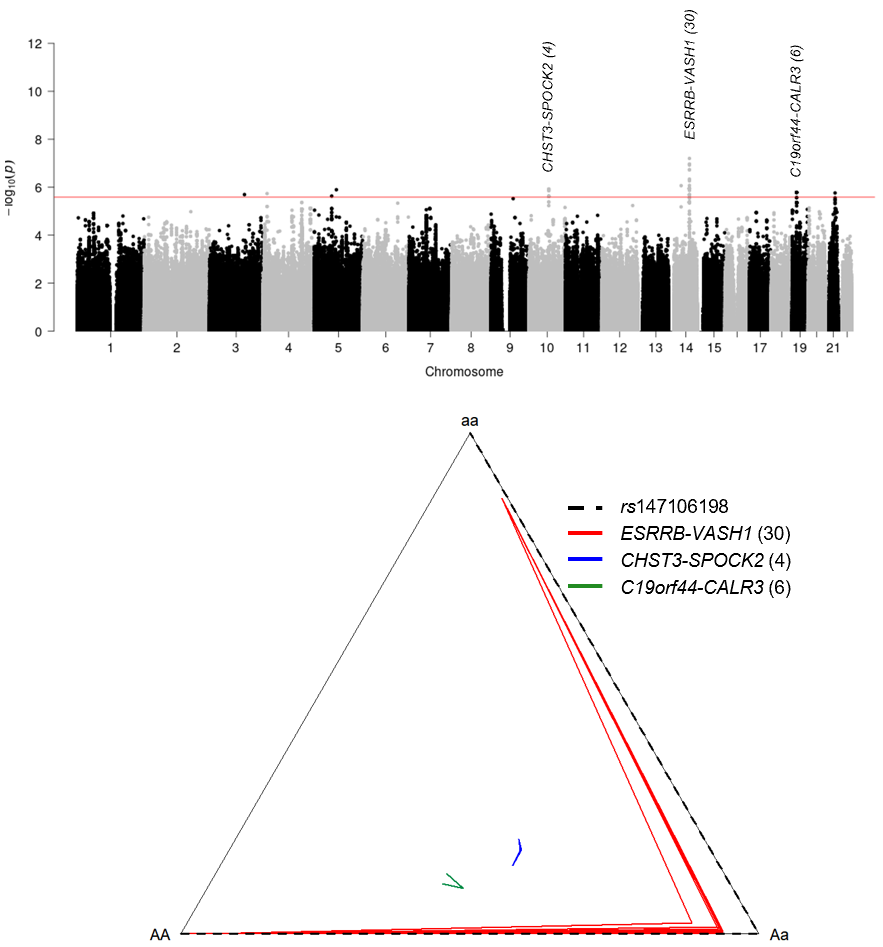}
\end{center}
\caption{\textbf{Upper:} Genome-wide CE4-Weibull analysis result; \textbf{Lower:} SNP cross-talk plot for 40 identified SNPs in relationship with the most top SNP $rs147106198$.}
\label{fig:manhattan}
\end{figure}

We first evaluated the treatment effect of ``AREDS formula" on time-to-progression in the overall population using a Weibull regression model, adjusting for the known risk factors including age, smoking status, and baseline severity score. The estimated HR between $Rx$ and $C$ is 1.15 with $p=0.26$, and the ratio of 75$th$ quantile progression-free time for $Rx$ and $C$ is 0.91 with $p=0.12$. It suggests that the combination of antioxidants and zinc does not seem to be effective in slowing down the disease progression in the overall population, which is consistent with previous findings \citep{prediction_genetics}. Then we applied CE4-Weibull method to analyze all common variants (i.e., MAF $\geq$ 0.05) across 22 autosomal chromosomes, resulting in a total of 3,837,556 SNPs. Similarly, baseline age, smoking status and severity score were adjusted. The upper panel of Figure~\ref{fig:manhattan} presents the Manhattan plot of this genome-wide CE4-Weibull analysis. By setting $m=10$, a total of 46 SNPs meet the significance threshold of $2.61\times 10^{-6} (= m/K)$. These SNPs are from nine gene regions on seven chromosomes. Following the recommendation rule we proposed in Section \ref{sec:100run}, there are three gene regions each with at least four SNPs meeting the $p$-value threshold and they are labeled in the Manhattan plot: \textit{CHST3-SPOCK2} on CHR 10 (4 SNPs), \textit{ESRRB-VASH1} on CHR 14 (30 SNPs), and \textit{C19orf44-CALR3} on CHR 19 (6 SNPs). We examined the correlation between all 46 identified SNPs using the cross-talk plot and presented the result in the lower panel of Figure~\ref{fig:manhattan}. We picked $rs147106198$ (from \textit{ESRRB-VASH1} region on CHR14), which has the smallest \textit{p}-value ($=7.00 \times 10^{-8}$) as the reference SNP. It can be seen that the other 29 SNPs from the same \textit{ESRRB-VASH1} region are highly correlated with the top SNP $rs147106198$, indicated by the long edges of the red segments. The other two gene regions, \textit{CHST3-SPOCK2} and \textit{C19orf44-CALR3} are not highly correlated with the \textit{ESRRB-VASH1} region, although the multiple SNPs within each region are highly dependent on each other (denoted by overlapped segments in green or blue color). In this case, there may be more than one causal SNP that leads to the differential treatment effects. 

\begin{table}[]
\caption {Baseline characteristics of the AREDS data}
\label{tab:tableone}
\scalebox{0.85}{
\begin{tabular}{lllll}
\hline
Number of subjects     & \begin{tabular}{@{}l@{}}All \\ (n=1170)\end{tabular}     & \begin{tabular}{@{}l@{}}Placebo \\ (n=754)\end{tabular}& \begin{tabular}{@{}l@{}}Antioxidants and Zinc \\ (n=416)\end{tabular} & \textit{p}-value* \\
Age                                                                           &                  &                  &                     & 0.309  \\
\multicolumn{1}{c}{Mean (SD)} & 68.4 (4.9)       & 68.3 (4.8)       & 68.6 (4.9)          &         \\
\multicolumn{1}{c}{Median (Range)} & 68.2 (55.3-81.0) & 68.0 (55.3-81.0) & 68.7 (55.5-79.5)    &          \\
Sex (n, \%)                                                                   &                  &                  &                     & 0.289       \\
\multicolumn{1}{c}{Female} & 655 (56.0)       & 413 (54.8)       & 242 (58.2)          &         \\
\multicolumn{1}{c}{Male} & 515 (44.0)       & 341 (45.2)       & 174 (41.8)          &         \\
Smoking (n, \%)                                                               &                  &                  &                     & 0.758       \\
\multicolumn{1}{c}{Never Smoked} & 571 (48.8)       & 371 (49.2)       & 200 (48.1)          &         \\
\multicolumn{1}{c}{Former/Current Smoker} & 599 (51.2)       & 383 (50.8)       & 216 (51.9)          &         \\
Baseline AMD severity score &                  &                  &                     & $<$0.001  \\
\multicolumn{1}{c}{Mean (SD)} & 3.2 (2.2)        & 2.6 (2.2)        & 4.0 (2.1)           &         \\
\multicolumn{1}{c}{Median (Range)} & 2.0 (1.0-8.0)    & 1.0 (1.0-8.0)    & 4.0 (1.0-8.0)       &         \\
Status (n, \%)                                                               &                  &                  &                     & $<$0.001       \\
\multicolumn{1}{c}{Progressed} & 269 (23.0)       & 133 (17.6)       & 136 (32.7)          &         \\
\hline
\multicolumn{5}{l}{{\small *\textit{p}-value is based on two-sample t test or Pearson Chi-square test for continuous or categorical variables }}\\
\end{tabular}}
\end{table}

\begin{figure}
\begin{center}
\includegraphics[width=6.5 in]{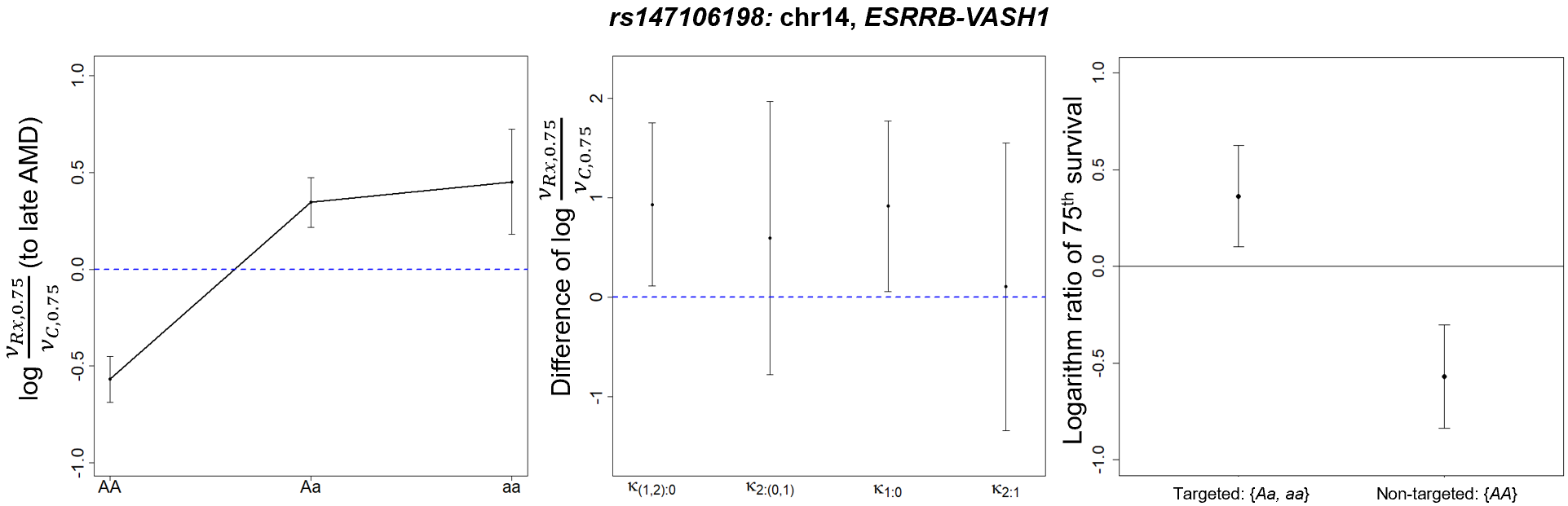}
\end{center}
\caption{Top identified SNP from AREDS, $rs147106198$. \textbf{Left:} treatment profile using log of the ratio of 75$th$ quantile survivals; \textbf{Middle:} CE4 results by taking the difference between the log ratio of 75$th$ quantile progression time to late AMD in the presented contrasts; \textbf{Right:} estimated treatment efficacy (log ratio of 75$th$ quantile survivals) in targeted and non-targeted groups.}
\label{fig:topSNP}
\end{figure}

We picked the top SNP ($rs147106198$ on CHR14) as our candidate marker for further discussion. Figure~\ref{fig:topSNP} demonstrates the treatment effect profiles and simultaneous confidence intervals for $rs147106198$, where the efficacy profile may suggest a dominant beneficial effect of \textit{a}. The CE4 simultaneous confidence intervals confirm that the targeted group is \textit{\{Aa,aa\}} combined since the confidence intervals of  $\kappa_{(1,2):0}$ and $\kappa_{1:0}$ are above the zero line. This targeted group consists about 52\% of the total patients, a reasonably high proportion of the entire population. Moreover, the estimated ratio of 75$th$ quantile progression-free times in the targeted and non-targeted groups (between $Rx$ and $C$) can be obtained, which are 1.44 and 0.57 for \textit{\{Aa,aa\}} and \textit{\{AA\}}, respectively, indicating that the combination of antioxidants and zinc treatment extends the progression time for 44\% compared to the placebo in the targeted group. The corresponding simultaneous 95\% confidence intervals for treatment efficacy (on the log scale) for the targeted and non-targeted population were also constructed and plotted in the right panel of Figure~\ref{fig:topSNP}. Finally, the characteristics of targeted and non-targeted population based on $rs147106198$ are summarized in Table~\ref{tab:charact_target}. As for the baseline characteristics and treatment assignment, the patients in the targeted group do not differ from the patients in the non-targeted group, indicating that the enhanced benefit from the treatment in the targeted population is plausibly due to the genetic difference rather than the demographic or clinical differences.

\begin{table}[]
\caption{Characteristics of targeted and non-targeted populations}
\label{tab:charact_target}
\scalebox{0.90}{
\begin{tabular}{lcccc}
\hline
& \multicolumn{4}{c}{$rs147106198$: chr14, \textit{ESRRB-VASH1} region} \\
\cline{3-5}
     &             & \begin{tabular}[c]{@{}c@{}}Targeted \end{tabular}                                             & \begin{tabular}[c]{@{}c@{}}Non-targeted \end{tabular} & \textit{p}-value \\
\begin{tabular}[c]{@{}l@{}}\# of subjects (n,\%)\end{tabular} && 605 (51.7)&565 (48.3) \\
Treatment efficacy $\frac{\hat{\nu}_{\textit{Rx,0.75}}}{\hat{\nu}_{\textit{C,0.75}}}$ $\dagger$ (SE) &&1.44 (1.01) &0.57 (1.01) &$6.99 \times 10^{-8}$\textsuperscript{$\star$} \\
Age                &                                                             &                                                              &                       & 0.560\\
\multicolumn{1}{c}{\begin{tabular}[c]{@{}c@{}}Mean (SD)\end{tabular}}      && \begin{tabular}[c]{@{}c@{}}68.5 (4.8)\end{tabular}       & 68.4 (4.9)       \\
\multicolumn{1}{c}{\begin{tabular}[c]{@{}c@{}}Median (range)\end{tabular}} && \begin{tabular}[c]{@{}c@{}}68.2 (55.3-81.0)\end{tabular} & 68.2 (55.8-80.5)  \\
\begin{tabular}[c]{@{}l@{}}Sex (n, \%)\end{tabular}  &                        &                                                              &                       & 0.982 \\
\multicolumn{1}{c}{Female}                                                  && \begin{tabular}[c]{@{}c@{}}338 (55.9)\end{tabular}       & 317 (56.1)   \\
\multicolumn{1}{c}{Male}                                                    && \begin{tabular}[c]{@{}c@{}}267 (44.1)\end{tabular}       & 248 (43.9) \\
\begin{tabular}[c]{@{}l@{}}Smoking (n, \%)\end{tabular}                 &     &                                                              &                       & 0.169 \\
\multicolumn{1}{c}{\begin{tabular}[c]{@{}c@{}}Never Smoked\end{tabular}}   && \begin{tabular}[c]{@{}c@{}}283 (46.8)\end{tabular}       & 288 (51.0)  \\
\multicolumn{1}{c}{\begin{tabular}[c]{@{}c@{}}Former/Current Smoker\end{tabular}}    && \begin{tabular}[c]{@{}c@{}}322 (53.2)\end{tabular}       & 277 (49.0) \\
\begin{tabular}[c]{@{}l@{}}Treatment (n, \%)\end{tabular}                 &     &                                                              &                       & 0.510 \\
\multicolumn{1}{c}{\begin{tabular}[c]{@{}c@{}}Placebo\end{tabular}}   && \begin{tabular}[c]{@{}c@{}}384 (63.5)\end{tabular}       & 370 (65.5)  \\
\multicolumn{1}{c}{\begin{tabular}[c]{@{}c@{}}Antioxidant $+$ Zinc\end{tabular}}    && \begin{tabular}[c]{@{}c@{}}221 (36.5)\end{tabular}       & 195 (34.5) \\
\begin{tabular}[c]{@{}l@{}}Baseline AMD severity score\end{tabular}          &&                                                              &                       & 0.487 \\
\multicolumn{1}{c}{\begin{tabular}[c]{@{}c@{}}Mean (SD)\end{tabular}}     & & \begin{tabular}[c]{@{}c@{}}3.1 (2.2)\end{tabular}        & 3.2 (2.2) \\
\multicolumn{1}{c}{\begin{tabular}[c]{@{}c@{}}Median (range)\end{tabular}} && \begin{tabular}[c]{@{}c@{}}2.0 (1.0-8.0)\end{tabular}    & 3.0 (1.0-8.0) \\
\hline
\multicolumn{3}{l}{\footnotesize $\dagger$: $\hat{\nu}$ denotes the estimated quantile progression time}\\
\multicolumn{5}{l}{\scriptsize $\star$: \textit{p}-value is from the corresponding CE4 contrast when simultaneous type I error is controlled, without adjusting for cross-SNP multiplicity}\\
\end{tabular}}
\end{table}

To help elucidate the controversial findings regarding whether genetic polymorphisms of \textit{CFH} and \textit{ARMS2} alter the treatment efficacy of AREDS formula, we closely checked 6 SNPs from these two regions and their results are presented in Table~\ref{tab:sixSNPs}. Note that $rs412852$, $rs1061170$, and $rs3766405$ from \textit{CFH} and $rs10490924$ from \textit{ARMS2-HTRA1} have been previously investigated in \cite{Seddon2016_AREDS, PNAS_harvard, Texas_2018}. We also reported the SNPs with the smallest CE4-based $p$-value from each region, which are $rs7522681$ and $rs11200647$. None of the 6 SNPs meets the significance threshold of $2.61\times 10^{-6}$, although 3 SNPs from \textit{CFH} region meet the nominal level of $0.05$. We further investigated $rs412852$ from $CFH$ and it seems our CE4-Weibull result suggests the combination group $\{AA,Aa\}$ exhibiting better treatment efficacy compared to its complementary group $\{aa\}$, which is similar to the findings from \cite{Seddon2016_AREDS} and \cite{Texas_2018}. However, it is worthwhile to note that from our genomewide CE4 analysis, none of these SNPs ranked top (Table~\ref{tab:sixSNPs}). Therefore, with appropriate multiplicity adjustment, neither \textit{CFH} or \textit{ARMS2-HTRA1} region has SNPs showing significant association with treatment efficacy, which is consistent with the conclusion indicated by \cite{Emily_2015}.

\begin{table}[]
\caption{CE4 results of six selected SNPs from \textit{CFH} and \textit{ARMS2-HTRA1} regions}
\label{tab:sixSNPs}
\begin{center}
\begin{tabular}{llll}
\hline
Gene                          & SNP                 & $p$.CE4       &rank.CE4 \\
\hline
\multirow{4}{*}{\textit{CFH}} & $rs7522681$  & $3.74\times10^{-4}$ &3843\\
                              & $rs412852$   & $9.49\times10^{-4}$ &9021\\
                              & $rs1061170$  & $1.22\times10^{-2}$ &71651\\
                              & $rs3766405$  & 0.38                &1614412       \\
\hline
\multirow{2}{*}{\textit{ARMS2-HTRA1}}         & $rs11200647$ & $7.75\times10^{-2}$ &378198 \\
                              & $rs10490924$ & 0.77 &3059694\\
\hline
\end{tabular}
\end{center}
\end{table}

It should be noted that based on different SNPs, the suggested targeted population may vary. In this example, if a top SNP from \textit{CHST3-SPOCK2} on CHR 10 is considered as the biomarker (e.g., $rs1245576$), the targeted population is about 65.8\% of the total population, which overlaps with the targeted population indicated by $rs147106198$ by 67.1\%. Our CE4-Weibull method provides reliable and interpretable candidate targeted populations for consideration, while the final decision on which population to target involves many other considerations such as development of companion diagnostics, labeling, marketing, and reimbursement.

\subsection{Validation on AREDS2}
\begin{figure}
\begin{center}
\includegraphics[width=6 in]{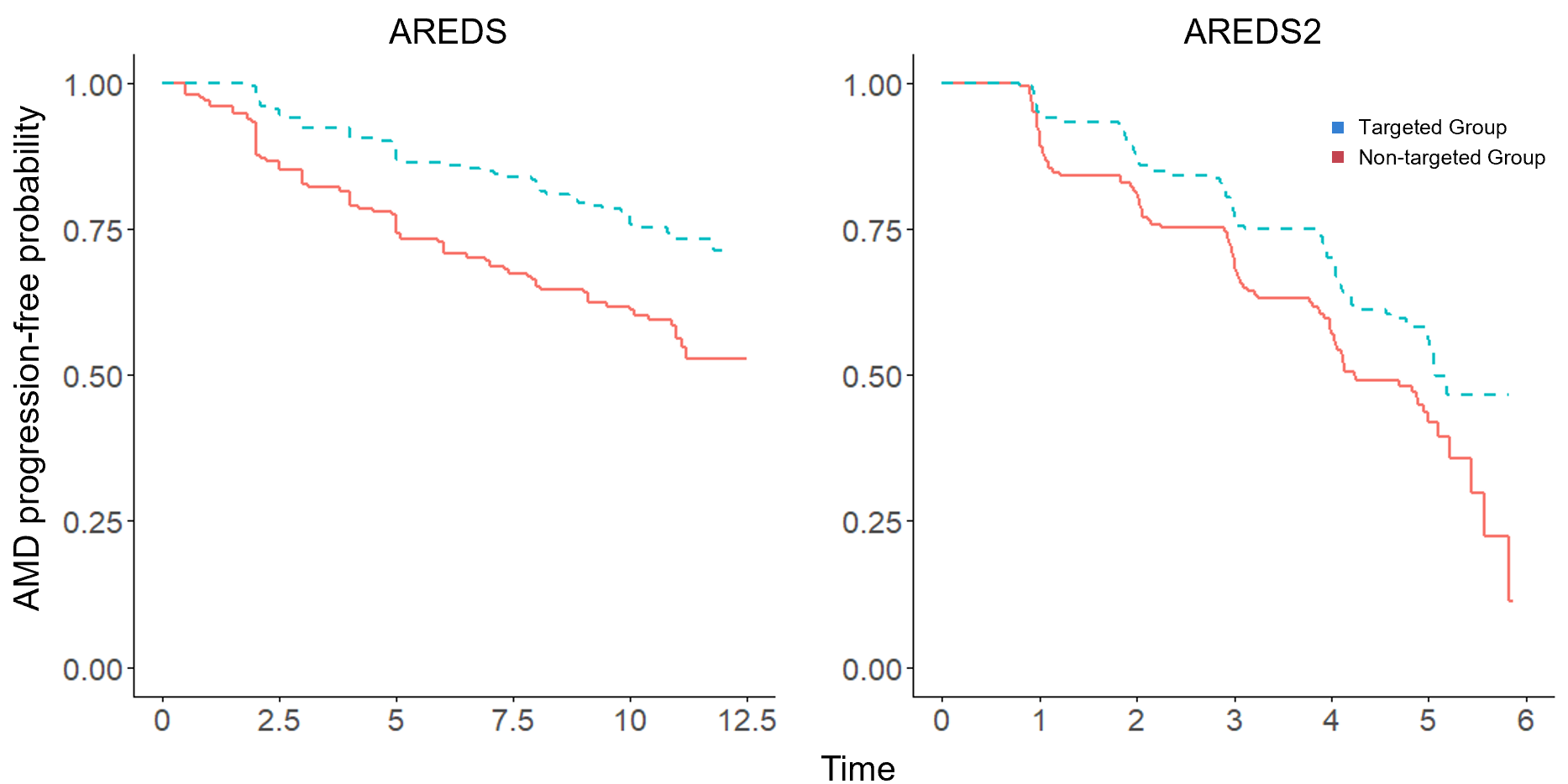}
\end{center}
\caption{Kaplan-Meier curves for targeted/non-targeted patients taking antioxidants and zinc combination in AREDS and AREDS2, where the subgroup is defined by $rs147106198$.\label{fig:validation}}
\end{figure}

AREDS2 was another independent large multi-center RCT of AMD \citep{AREDS2}. It was designed to evaluate the effect of refined AREDS formulations on AMD progression, as compared to the original AREDS formulation. Participants of AREDS2 were more severe at baseline and the follow-up time was only about half of the AREDS's follow-up time. Four arms were included: AREDS formulation, AREDS formulation plus Lutein/Zeaxanthin, AREDS formulation plus DHA/EPA, and AREDS formulation plus Lutein/Zeaxanthin and DHA/EPA. Since there is no real placebo arm in AREDS2, we cannot apply CE4-Weibull to identify subgroups with enhanced efficacy of AREDS formulation directly in AREDS2. Instead, we specifically investigated the patient's response to the same antioxidants and zinc combination treatment to check whether we observe similar differential response patterns between the AREDS identified targeted and non-targeted groups in AREDS2 as compared to AREDS. We used the same SNP $rs147160198$ to determine whether a patient belongs to targeted \textit{\{Aa,aa\}} group or non-targeted \textit{\{AA\}} group in both studies. Table~\ref{tab:AREDS_AREDS2} presents the patient characteristics within the targeted and non-targeted groups, separately for AREDS and AREDS2. None of the baseline risk factors differs between targeted and non-targeted populations in each study. Between the two studies, AREDS2 patients are older and more severe and thus are anticipated to progress faster. Figure~\ref{fig:validation} compares the progression-free Kaplan-Meier curves between the targeted and non-targeted groups within each study. In both studies, the targeted population shows an obvious better progression-free profile than the non-targeted population, with the log-rank test $p$value of 0.00011 and 0.013, respectively. Therefore, we successfully validated our identified targeted group by SNP $rs147160198$ in AREDS2. We also checked the subgroups indicated by reported SNPs from the \textit{CFH} region, for example $rs412852$. In AREDS, the two groups ($\{AA, Aa\}$ vs $aa$) exhibit differential treatment response profiles (log-rank $p<0.001$). However, such differential response profiles were not observed in AREDS2. This further emphasizes appropriate multiplicity adjustment is crucial for robust subgroup identification.

\begin{sidewaystable}
\caption{Characteristics of targeted and non-targeted populations in AREDS and AREDS2}
\label{tab:AREDS_AREDS2}
\begin{minipage}[t]{\textheight}
\scalebox{0.90}{
\begin{tabular}{lcccccccc}
\hline
                                                                                 & \multicolumn{4}{c}{AREDS}                                                                & \multicolumn{4}{c}{AREDS2}                                                                \\
\cline{3-5}
\cline{7-9}
     &             & \begin{tabular}[c]{@{}c@{}}Targeted \end{tabular}                                             & \begin{tabular}[c]{@{}c@{}}Non-targeted \end{tabular} & \textit{p}-value* && \begin{tabular}[c]{@{}c@{}}Targeted \end{tabular}                                             & \begin{tabular}[c]{@{}c@{}}Non-targeted \end{tabular} & \textit{p}-value* \\
\begin{tabular}[c]{@{}l@{}}\# of subjects (n,\%)\end{tabular} && 221 (53.1)&195 (46.9) &&&164 (51.1) &157 (48.9)\\
Age                &                                                             &                                                              &                       & 0.696   &                                                     & &                      & 0.242 \\
\multicolumn{1}{c}{\begin{tabular}[c]{@{}c@{}}Mean (SD)\end{tabular}}      && \begin{tabular}[c]{@{}c@{}}68.6 (4.8)\end{tabular}       & 68.8 (5.0)            &         & &\begin{tabular}[c]{@{}c@{}}70.2 (7.4)\end{tabular}       & 71.1 (7.8)\\
\multicolumn{1}{c}{\begin{tabular}[c]{@{}c@{}}Median (range)\end{tabular}} && \begin{tabular}[c]{@{}c@{}}68.4 (55.5-78.5)\end{tabular} & 68.9 (56.1-79.5)      &         && \begin{tabular}[c]{@{}c@{}}71.0 (51.0-85.0)\end{tabular} & 71.0 (53.0-86.0)\\
\begin{tabular}[c]{@{}l@{}}Sex (n, \%)\end{tabular}  &                        &                                                              &                       & 0.700    &                                                              &    &                  & 0.824 \\
\multicolumn{1}{c}{Female}                                                  && \begin{tabular}[c]{@{}c@{}}131 (59.3)\end{tabular}       & 111 (56.9)            &         & &\begin{tabular}[c]{@{}c@{}}96 (58.5)\end{tabular}       & 89 (56.7)\\
\multicolumn{1}{c}{Male}                                                    && \begin{tabular}[c]{@{}c@{}}90 (40.7)\end{tabular}       & 84 (43.1)            &         && \begin{tabular}[c]{@{}c@{}}68 (41.5)\end{tabular}       & 68 (43.3)\\
\begin{tabular}[c]{@{}l@{}}Smoking (n, \%)\end{tabular}                 &     &                                                              &                       & 0.806   &                                                              &    &                  & 0.587 \\
\multicolumn{1}{c}{\begin{tabular}[c]{@{}c@{}}Never Smoked\end{tabular}}   && \begin{tabular}[c]{@{}c@{}}108 (48.9)\end{tabular}       & 92 (47.2)            &         && \begin{tabular}[c]{@{}c@{}}77 (47.0)\end{tabular}       & 68 (43.3)\\
\multicolumn{1}{c}{\begin{tabular}[c]{@{}c@{}}Former/Current Smoker\end{tabular}}    && \begin{tabular}[c]{@{}c@{}}113 (51.1)\end{tabular}       & 103 (52.8)            &         && \begin{tabular}[c]{@{}c@{}}87 (53.0)\end{tabular}       & 89 (56.7)\\
\begin{tabular}[c]{@{}l@{}}Baseline AMD severity score\end{tabular}          &&                                                              &                       & 0.375   &                                                              & &                     & 0.347\\
\multicolumn{1}{c}{\begin{tabular}[c]{@{}c@{}}Mean (SD)\end{tabular}}     & & \begin{tabular}[c]{@{}c@{}}4.0 (2.1)\end{tabular}        & 4.2 (2.1)             &         && \begin{tabular}[c]{@{}c@{}}6.5 (1.1)\end{tabular}        & 6.6 (1.0)\\
\multicolumn{1}{c}{\begin{tabular}[c]{@{}c@{}}Median (range)\end{tabular}} && \begin{tabular}[c]{@{}c@{}}4.0 (1.0-8.0)\end{tabular}    & 4.0 (1.0-8.0)         &         && \begin{tabular}[c]{@{}c@{}}7.0 (2.0-8.0)\end{tabular}    & 7.0 (2.0-8.0)\\
\hline
\multicolumn{7}{l}{{\small *\textit{p}-value is based on two-sample t test or Pearson Chi-square test for continuous or categorical variables }}\\
\end{tabular}}
\end{minipage}
\end{sidewaystable}

\subsection{Data Availability}
The phenotype and genotype data of AREDS and AREDS2 are available from the online repository dbGap (accession: $phs000001:v3:p1$, and $phs001039:v1:p1$, respectively). The proposed method and its applications are implemented in R. The key functions can be obtained from GitHub upon the acceptance of this manuscript.

\section{Conclusion and Discussion}
\label{sec:conclusion}
In this work, we develop a new statistical method to confidently identify and infer subgroups in modern RCTs with time-to-event outcomes. Different from machine learning based approaches, our CE4-Weibull approach, derived from the fundamental multiple testing principle, provides simultaneous confidence intervals on contrasts that directly compare the treatment efficacy between subgroups or combination of subgroups. The contrasts are built upon a logic-respecting efficacy measure, the ratio of quantile survival times (between $Rx$ and $C$), which enjoys the unique covariate invariant property in addition to its interpretation-friendly feature. 

Our CE4-Weibull adjusts for multiplicity both within and across the markers. It rigorously combines two error rate controls, familywise error rate control within each marker, and per family error rate control across the markers. Such error control is appropriate in a drug development process, as it allows flexibility in the exploration of multiple candidate markers, while being confident in the patient subgroup to target from any selected marker(s). Such a novel and rigorous multiplicity adjustment contributes to the reduction in the so-called ``reproducibility crisis'' in which many discoveries in markers or effective subgroups turn out to be false positive findings.

From our realistic simulation studies where the SNP data are taken from true individuals, we recommend practically useful rules for identifying ``candidate" markers. Finally, we successfully applied our method on AREDS data to identify subgroups that exhibit enhanced treatment efficacy with combination of antioxidant supplements and zinc in delaying AMD progression. After conducting the genomewide analysis using our CE4-Weibull method, three gene regions were discovered to suggest subgroups with significantly enhanced efficacy: \textit{CHST3-SPOCK2} on CHR 10, \textit{ESRRB-VASH1} on CHR 14, and \textit{C19orf44-CALR3} on CHR 19. We further validated the subgroups defined by the top SNP $rs147106198$ from \textit{ESRRB-VASH1} region using the data from an independent AREDS2 study. \cite{ESRRB} found that the estrogen related receptor beta (\textit{ESRRB}) was (weakly) associated with CNV AMD. \cite{VASH1} first demonstrated the angiogenesis modulation of \textit{VASH} is involved in the pathological process of AMD. Later \cite{Zeng_2012} inferred that the treatment with AREDS formulation is likely to affect both angiogenesis and endothelial-macrophage interactions. Thus our findings provide new perspectives on the differential treatment efficacy, suggested by genetic polymorphisms for delaying AMD progression.

Although we use the SNP testing scenario to demonstrate our method, the key elements of the method apply to broader scenarios with all kinds of markers to consider. When the marker separates the patient population into more groups ($>3$), additional contrasts need to be considered to obtain the complete ordering of the treatment efficacy, which will then be used to identify the subgroups. The current version of our method only handles discrete markers and more work is required to generalize it for continuous markers. In doing that, one may borrow the idea from \cite{thresholding} which considers all candidate thresholds for a continuous marker when deriving simultaneous confidence intervals.

\bibliographystyle{apalike}

\bibliography{bibCE4}
\end{document}